\documentclass[aps,prl,amssymb,twocolumn]{revtex4-1}%
\usepackage{graphicx}
\usepackage{dcolumn}
\usepackage[T2A]{fontenc}
\usepackage[cp1251]{inputenc}

\begin{document}

\title{Mooij Rule and Weak Localization}

\author{V. F. Gantmakher}

\affiliation{\mbox{Institute of Solid State Physics, Russian Academy of Sciences, Chernogolovka, Moscow region, 142432 Russia}}

\begin{abstract}
It has been shown that the observed correlation between the resistivity $\rho$ of high-resistive metallic alloys and the sign of the temperature derivative $d\rho/dT$ can be explained by taking into account the weak localization. This correlation is known as Mooij rule: the derivative $d\rho/dT$ is negative for alloys with resistivity in the range of $300\div150\,\mu\Omega\cdot$cm, which corresponds to the electron mean free path  about the interatomic distance; however, this derivative is positive for alloys with lower resistivity.
\end{abstract}

\maketitle

This paper considers metals and alloys where a carrier concentration $n$ is
not lower than one electron per atom, i.e., $n\gtrsim a^{-3}$ (here $a$ is the interatomic distance). The Anderson transition is never observed in these materials; i.e., they remain metallic under any disorder level with the resistivity lower than the critical value $\rho^* \sim300\,\mu\Omega\cdot$cm, which is approximately equal to the resistivity at the mean free path of electrons $l$ about the interatomic distance, i.e.,
\begin{equation}\label{R*}
  \rho^*=\frac{\hbar k_F}{ne^2l}=
  \frac{\hbar}{e^2}\frac{1}{k_F} \qquad l\sim k_F^{-1}\sim a.
\end{equation}
Here, $k_F$ is the Fermi wavenumber and the frequency $\hbar/\tau$ of electron scattering by static defects is about the Fermi energy $T_F$ (temperature is given everywhere in energy units).

The absence of the Anderson transition in metallic materials with a high electron concentration has been confirmed in three types of experiments: \par\noindent
(i) measurements of the resistance of these materials at low temperatures under increased disorder,  \par\noindent
(ii) measurements of the temperature dependence of the resistance of high-resistive alloys [1],  \par\noindent
(iii) measurements of the temperature dependence of the resistance of alloys with a comparatively low residual resistivity $\rho_0$, but with a very large electron–phonon coupling constant, i.e., with the resistance rapidly increasing with the temperature [2].

This work is devoted to the second of the three listed approaches.

Let the level of disorder be characterized by the normalized frequency of scattering by static defects given by the expression
\begin{equation}\label{alpha}
  \alpha=\frac{\hbar}{T_F}\tau^{-1} , \qquad 0\leqslant\alpha\leqslant1.
\end{equation}
According to the Ioffe–Regel criterion, $\alpha$ cannot be larger than unity. The classical residual resistivity at $\alpha=1$ is $\rho_0\approx \rho^*$. The value $\alpha=0$ corresponds to an ideal defect-free lattice in which $\rho_0=0$.

At $\alpha\ll1$, scattering by disorder is independent of scattering by phonons and the temperature-dependent part of the resistivity, which is called the Gruneisen function $G(T)$, is independent of disorder, so that the total resistivity is $\rho_{cl}=\rho_0+G(T)$ (subscript 'cl' indicates that interference corrections to the resistivity due to the wave nature of the electron are ignored). For convenient comparison with Eq. (1), it is reasonable to represent $G(T)$ in the form $G(T)=(\hbar/e^2)(\alpha_{ph}(T)/k_F)$, so that
\begin{equation}\label{Rcl}
 \rho_{cl}=\rho_0+G(T)
 =\frac{\hbar}{e^2}\frac{1}{k_F}(\alpha+\alpha_{ph}).
\end{equation}
At low temperatures $T\rightarrow0$ the function $G(T)$ behaves as $\alpha_{ph}(T)\propto T^5$, and at $T\gtrsim T_D/3$ ($T_D$ is the Debye temperature), this function is linear [3]:
\begin{equation}\label{G(T)}
\alpha_{ph}(T)\rightarrow(\gamma T/T_F),
\end{equation}
where the numerical coefficient $\gamma$ depends on the properties of a particular material. In the temperature range, where asymptotic expression (4) is valid, $\alpha_{ph}$ is related to the electron–phonon scattering frequency $\tau_{ph}^{-1}$ through the following formula similar to Eq. (2):
\begin{equation}\label{alpha-ph}
  \alpha_{ph}=\frac{\hbar}{T_F}\tau_{ph}^{-1}=\gamma T/T_F .
\end{equation}

According to estimates, where the deformation potential is taken to be $D\simeq e^2/a$ and $k_F$ is equal to is reciprocal lattice vector $K$, the coefficient $\gamma=1$ [4]. In real metals, $\gamma$ can be both smaller and larger unity. The condition $\gamma\gg1$ means strong electron–phonon coupling.

Since the thermal velocity of ions is much lower than the velocity of electrons $v_F$, static disorder from lattice defects and phonon-induced dynamic disorder provide the same action on electrons. For the classical resistivity given by Eq. (3) to satisfy the Ioffe–Regel criterion, it is necessary that
$$\alpha+\alpha_{ph}\leqslant1.$$

In high-resistive alloys with $\alpha\gtrsim0.3\div0.4$, Eq. (3) is violated because the static and dynamic parts of the resistivity cease to be independent. The temperature dependent part of the resistivity becomes much smaller and even may have the opposite sign. Numerous experimental observations were summarized by Mooij [1], who formulated the following empirical rule. The sign of the derivative of the resistivity $d\rho/dT$ of high-resistive alloys correlates with the resistivity: \mbox{$d\rho/dT>0$} in metallic alloys with the resistivity lower than $100\div150\,\mu\Omega\cdot$cm, but $d\rho/dT< 0$ in alloys with a higher resistivity; i.e.,
\begin{equation}\label{RoolM}
\left.
 \begin{array}{lc}
  d\rho/dT>0 \qquad\mbox{at}\quad \rho< \\
  d\rho/dT<0 \qquad\mbox{at}\quad \rho> \
 \end{array}
\right\}
 100\div150\,\mu\Omega\cdot\mbox{cm}
\end{equation}
(see also Chapter 1 in [5]). This is valid for comparatively high temperatures from 20–30 K to room temperature or even higher.

The first attempt to theoretically explain the Mooij rule was made in [6] using the scaling theory [7] of the metal–insulator quantum transition. However, this transition is not observed in these systems and the assumption that these systems are near this transition
is not convincing. The remark in [8, Sect. 2.5.3] that the Mooij rule can be attributed to weak localization, which is responsible for the negative temperature coefficient of resistivity, seems more interesting. The possibility of explaining empirical dependence (6) by taking into account weak localization in the simple model is analyzed in this work developing the remark made in [8].

For definiteness, let the melting temperature of the metallic alloy $T_{\rm melt}$ and Debye temperature  $ T_D$ satisfy the relations
\begin{equation}\label{Tmelt}
 T_{\rm melt}\simeq T_F/10, \qquad T_D\simeq T_{\rm melt}/3.
\end{equation}
Taking into account asymptotic expression (4), this means that  $\alpha_{ph}$ is proportional to $T$ in the interval
\begin{equation}\label{Interval}
0.01\leqslant\alpha_{ph}\leqslant0.1,
\end{equation}
which is of main interest on the  [$\alpha_{ph},\alpha$]-plane. Interval (8) is
marked in Fig.\,1 by two vertical straight segments. In order to expand this interval, the logarithmic scale in the $\alpha_{ph}$ axis is used. In the region above the lower line $\alpha=\alpha_{ph}\; (\gamma=1)$ scattering by static defects occurs more frequently than scattering by phonons. In particular, for high-resistive alloys ($\alpha>0.3$) in the temperature range $T/T_F<0.1\div0.2$ the following inequality is valid:
\begin{equation}\label{alpha-inequ}
 \alpha\gg\alpha_{ph}.
\end{equation}
For this reason, the diffraction of an electron wave on a random potential field of impurities in the case of the high static scattering frequency $\tau^{-1}$ should occur in this range up to high temperatures. At  $\alpha\ll1$, when the electron wavefunction is a long wave train and calculations can be performed in the first approximation of perturbation theory, diffraction results in the weak localization effect [9-11]. The quantum correction  $\delta\sigma$ to the conductivity appears; it is limited by phonon scattering and is expressed in terms of the diffusion length  $L_\varphi$:\begin{equation}\label{Lphi}
  \delta\sigma=\frac{e^2}{\hbar}\frac{1}{L_\varphi}=
  \frac{e^2}{\hbar}k_F(\alpha\alpha_{ph})^{1/2},\quad
  L_\varphi=l\sqrt{\tau_{ph}/\tau}.\,
\end{equation}
Strictly speaking, in the region $\alpha\sim1$, it is impossible to use only the first approximation and Eqs. (10) of weak localization. However, it is instructive to analyze
how the relative contribution of this first approximation changes with an increase in $\alpha$.

\begin{figure}[t]
\includegraphics{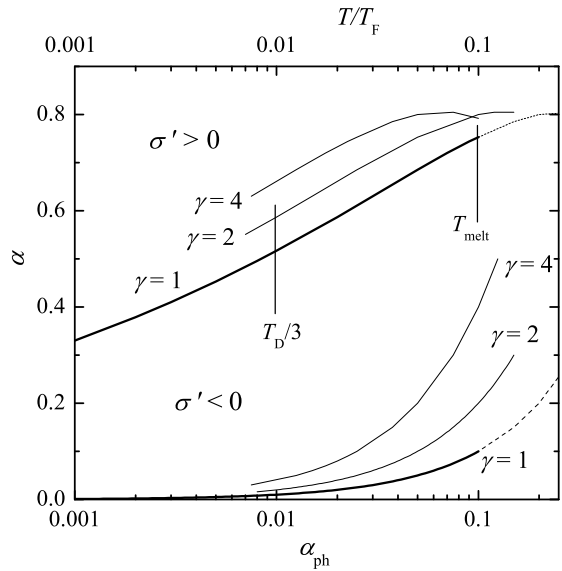}
\caption{Plane [$\alpha_{ph},\alpha$]. Along three lower lines $\tau^{-1}=\tau^{-1}_{ph}$ (at different $\gamma$). Along three upper lines the temperature derivative of the conductivity is zero, i.e., $\partial\sigma/\partial T=0$, also at different $\gamma$.  }
\end{figure}

The addition of quantum correction (10) to the classical conductivity $\sigma_{cl}=1/\rho_{cl}$ for the three-dimensional medium yields \begin{equation}\label{sigma}
  \sigma=\sigma_{cl}+\delta\sigma=
  \frac{e^2}{\hbar}k_F\left[\frac{1}{\alpha+\alpha_{ph}}+(\alpha\alpha_{ph})^{1/2}\right].
\end{equation}

Here, the quantity $\alpha_{ph}$ which increases monotonically with the temperature, appears in the denominator of the first term and the numerator of the second term. The condition of zero derivative $\partial\sigma/\partial\alpha_{ph}$ gives the equation
\begin{equation}\label{deriv}
  \alpha+\alpha_{ph}=\sqrt{2}(\alpha/\alpha_{ph})^{1/4}
\end{equation}
of the line along which this derivative changes sign (the upper line $\gamma=1$ in Fig.\,1).

The assumption that  $\gamma=1$ means that not only the lower scale $\alpha_{ph}$, but also the upper scale $T/T_F$ can be used for the corresponding lines in Fig.\,1. At $\gamma\neq1$, only the upper scale should be used after the corresponding lines are shifted by $\lg\gamma$ in the horizontal direction. The weak localization regime occurs above the corresponding lower line and the temperature derivative of the conductivity  $\sigma'>0$ is positive (i.e., $\partial \rho/\partial T$ is negative) above the corresponding upper line.

The resulting set of lines indicates that the temperature derivative of the conductivity for any $\gamma$ value is positive if  $\alpha>0.8$ i.e., the contribution from weak localization dominates; this derivative is negative for $\alpha<0.5$, at least in temperature range (8). This is the Mooij rule.

In the expressions for the classical conductivity $\sigma$ and the quantum correction $\delta\sigma$ written above, the electron-electron scattering was not taken into account. Indeed, in diffusion regime (9), the electron-electron collision frequency $\tau_{ee}$ in the three dimensional system is determined by the expression [11] (see also Sect. 2.4 in [5])
\begin{equation}\label{tauee}
\hbar/\tau_{ee}\sim T^{3/2}T_F^{-2}(\hbar/\tau)^{3/2}.
\end{equation}
At strong disorder $(\alpha\approx1)$, this expression is modified into the form
\begin{equation}\label{tauee1}
\hbar/\tau_{ee}\sim T(T/T_F)^{1/2}\ll\hbar/\tau_{ph}.
\end{equation}
According to this relation, the direct contribution from the electron–electron scattering to the classical resistivity given by Eq. (3) and to the classical conductivity in temperature range (8) can be neglected.

On the other hand, the electron–electron interaction also produces the quantum correction to the conductivity [11]. According to the estimates made in [3], the Altshuler–Aronov quantum correction in three-dimensional systems is larger than the weak localization correction; it is often observed in high-resistive alloys [12, 13]. This should increase the part of Eq. (11) with positive derivative.

To summarize, the experimental temperature dependence of the resistivity of high-resistive alloys has been described taking into account quantum corrections obtained in the first approximation of perturbation theory. This may mean that the contribution
from higher orders is small for some reasons in agreement with the absence of the transition from weak localization to strong localization (Anderson transition) in these alloys. Localization remains weak; however, due to a very short mean free path $l\sim k_F^{-1}$, weak localization holds up to high temperatures and is responsible for the Mooij rule.

I am grateful to Yu.M. Gal’perin, A.S. Ioselevich, and V.P. Mineev for valuable remarks. This work was supported by the Russian Foundation for Basic Research, project no. 11-02-12071.

\end{document}